\documentclass{PoS}
\usepackage{float}
\usepackage{graphicx}
\usepackage{caption}

\title{A positioning system for Baikal-GVD}

\ShortTitle{Baikal-GVD positioning}

\author{\speaker{A.D.~Avrorin}$^{,a}$, A.V.~Avrorin$^a$, V.M.~Aynutdinov$^a$, R.~Bannash$^g$, I.A~Belolaptikov$^b$, V.B.~Brudanin$^b$, N.M.~Budnev$^c$, G.V.~Domogatsky$^a$, A.A.~Doroshenko$^a$, R.~Dvornick\'y$^{b,h}$, A.N.~Dyachok$^c$, Zh.-A.M.~Dzhilkibaev$^a$, L. Fajt$^{b,h,i}$, S.V~Fialkovsky$^e$, A.R.~Gafarov$^c$, K.V.~Golubkov$^a$, N.S.~Gorshkov$^b$, T.I.~Gress$^c$, R.~Ivanov$^b$, K.G.~Kebkal$^g$, O.G.~Kebkal$^g$, E.V.~Khramov$^b$ , M.M.~Kolbin$^b$, K.V.~Konischev$^b$, A.V.~Korobchenko$^b$, A.P.~Koshechkin$^a$, A.V.~Kozhin$^d$, M.V.~ Kruglov$^b$, M.K.~Kryukov$^a$, V.F.~Kulepov$^e$, M.B.~Milenin$^a$, R.A.~Mirgazov$^c$, V.~Nazari$^b$, \fbox{A.I.~Panfilov$^a$}, D.P.~Petukhov$^a$ E.N.~Pliskovsky$^b$, M.I.~Rozanov$^f$, E.V.~Rjabov$^c$, V.D.~ Rushay$^b$, G.B.~Safronov$^b$, B.A.~Shaybonov$^b$, M.D.~Shelepov$^a$, F.~\u{S}imkovic$^{b,h,i}$, A.V.~Skurikhin$^d$, A.G.~Solovjev$^b$, M.N.~ Sorokovikov$^b$, I.~\u{S}tekl$^i$, E.O.~Sushenok$^b$, O.V.~Suvorova$^a$, V.A.~Tabolenko$^c$, B.A.~Tarashansky$^c$, and S.A.~Yakovlev$^g$\\
$^a$ Institute for Nuclear Research, Russian Academy of Sciences, Moscow, 117312 Russia\\
$^b$ Joint Institute for Nuclear Research, Dubna, 141980 Russia\\
$^c$ Irkutsk State University, Irkutsk, 664003 Russia\\
$^d$ Institute of Nuclear Physics, Moscow State University, Moscow, 119991 Russia\\
$^e$ Nizhni Novgorod State Technical University, Nizhni Novgorod, 603950 Russia\\
$^f$ St. Petersburg State Marine Technical University, St. Petersburg, 190008 Russia\\
$^g$ EvoLogics Gmbh, Germany\\ 
$^h$ Comenius University, Mlynska Dolina F1, Bratislava, 842 48 Slovakia\\
$^i$ Czech Technical University in Prague, Prague, 128 00 Czech Republic\\
E-mail: \email{avrorin@inr.ru}
}

\abstract{
A cubic kilometer scale neutrino telescope Baikal-GVD is currently under construction in Lake Baikal. 
Baikal-GVD is designed to detect Cerenkov radiation from products of astrophysical neutrino interactions with Baikal water by a lattice of photodetectors submerged between the depths of 1275 and 730 m.
The detector components are mounted on flexible strings and can drift from their initial positions upwards to tens of meters.
This introduces positioning uncertainty which translates into a timing error for Cerenkov signal registration.
A spatial positioning system has been developed to resolve this issue.
In this contribution, we present the status of this system, results of acoustic measurements and an estimate of positioning error for an individual component.
}

\FullConference{36th International Cosmic Ray Conference -ICRC2019-\\
		July 24th - August 1st, 2019\\
		Madison, WI, U.S.A.}

\begin{document}

\section{Introduction}
Baikal-GVD \cite{status} is a deep underwater neutrino telescope currently under construction in Lake Baikal. 
It is designed to measure the direction and energy of astrophysical neutrinos by detecting Cherenkov radiation emitted by particles produced in neutrino interactions with water with an array of spatially distributed photomultipliers.
Each photomultiplier, along with a controller and calibration equipment is placed in a transparent spherical pressure housing with a diameter of 42 cm, comprising an optical module (OM). 
OMs are installed on strings - flexible cables strung between an anchor and subsurface buoys.
Each string has 36 OMs installed at 15 meter intervals starting at the depth of 1270 meters.
The strings are grouped into clusters, each cluster having an independent trigger system.
A cluster consists of 8 strings installed in the vertices and center of a regular heptagon with a radius of 60 meters.
Some clusters also possess an instrumentation string used for calibration and hardware tests.
Following the installation of clusters 4 and 5 during the winter expedition 2019, a total of 5 clusters have been deployed.

\begin{figure}
  \center
  \begin{minipage}{0.65\linewidth}
    \center
    \includegraphics[height=5cm]{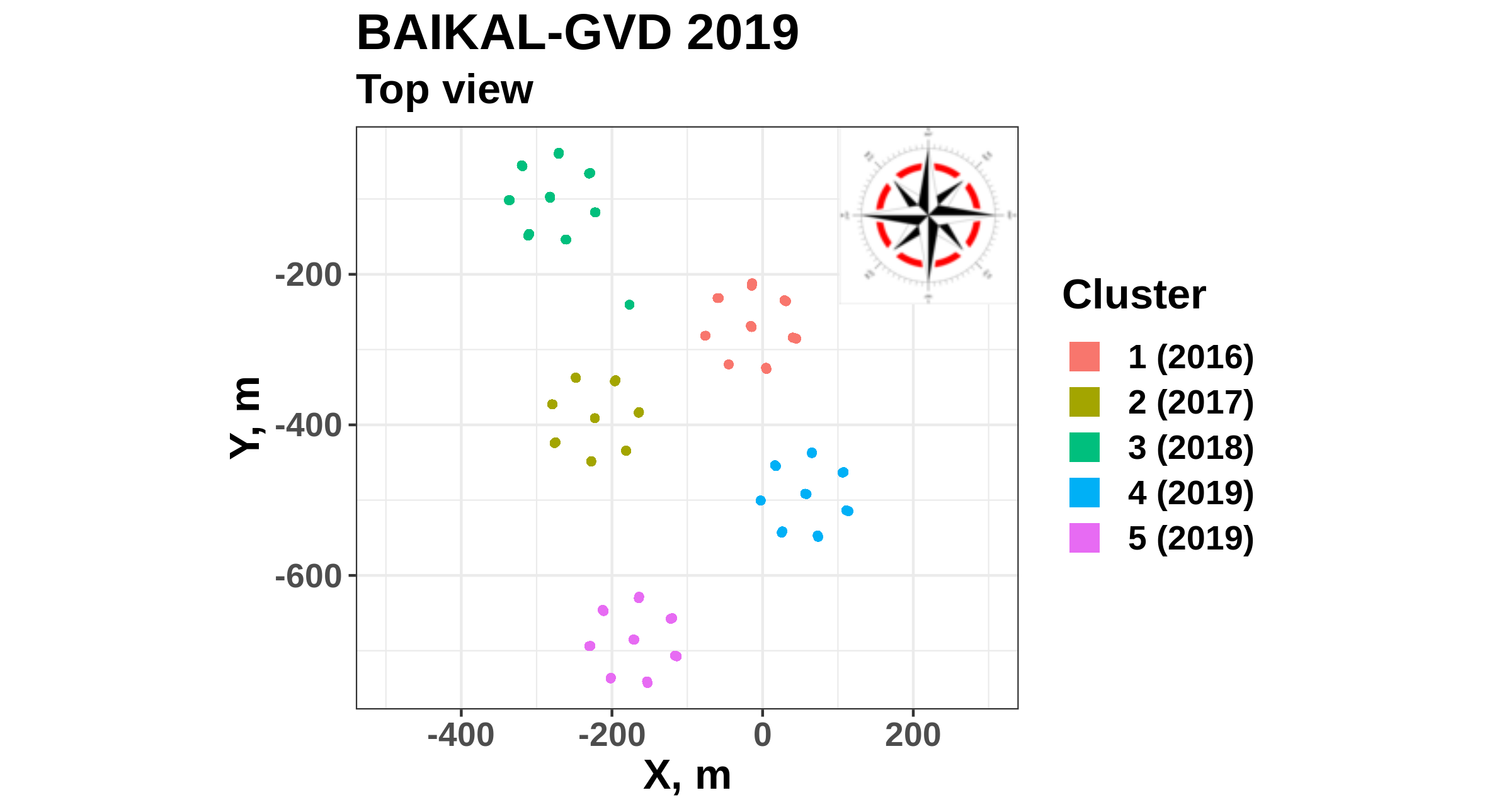}
    \caption{Planar beacon coordinates, 2019}
    \label{xy-top}
  \end{minipage}
  \begin{minipage}{0.32\linewidth}
    \center
    \includegraphics[height=5cm]{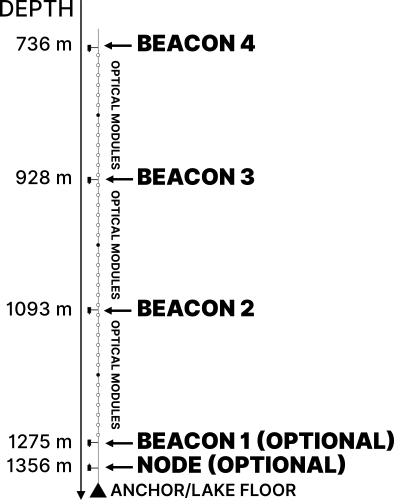}
    \caption{AM layout on a string}
    \label{am-layout}
  \end{minipage}
\end{figure}

Due to string flexibility and currents in Lake Baikal, OMs can drift beyond 50 meters from their median positions.
Uncertainty in OM coordinates introduces a time calibration error and so monitoring OM coordinates is crucial for event reconstruction. 

To solve this problem, Baikal-GVD uses a hydroacoustic positioning system (APS) \cite{aps2011}.
APS consists of a network of EvoLogics S2C R42/65 acoustic modems (AMs) installed on the strings of the detector.
As shown on Figure \ref{am-layout}, a baseline acoustic configuration consists of three beacons (downward oriented AMs) installed along OMs and either a bottom beacon or a node (an upward oriented AM).
In the 2019 configuration, Baikal-GVD has a total of 171 AMs, 22 of them - nodes.

The nodes are attached at the base of a string, near the anchor, and are considered stationary.
The node coordinates are determined shortly after the string installation is completed.
An AM is submerged at the depth of $\sim1$ meter at several sites on  the surface and the acoustic distances to the nodes are measured.
The node positions are then trilaterated from the acquired distances and GPS coordinates of each site.

Beacon coordinates are acquired by measuring acoustic distances to the antenna formed by the nodes on the lake floor.
Once the acoustic distances are measured, each beacon is trilaterated from the known node coordinates.
The communication between AMs is facilitated by the D-MAC protocol \cite{dmac}.
The planar beacon coordinates for the 2019 configuration of Baikal-GVD are shown on Figure \ref{xy-top}.

Beacon coordinates are reconstructed online. 
At regular intervals ($\sim1$ minute per cluster) the APS software on the shore polls the beacons for the acoustic distances, then performs the reconstruction. 
After reconstruction, beacon coordinates are buffered at the shore and then transferred to the data storage in JINR.
OM coordinates are acquired by interpolating beacon coordinates, presuming a piece-wise linear model of the string.
If necessary, coordinates of other components installed on the string are acquired the same way.

\section{Performance}

\begin{figure}
  \begin{minipage}{0.32\linewidth}
    \includegraphics[width=\linewidth]{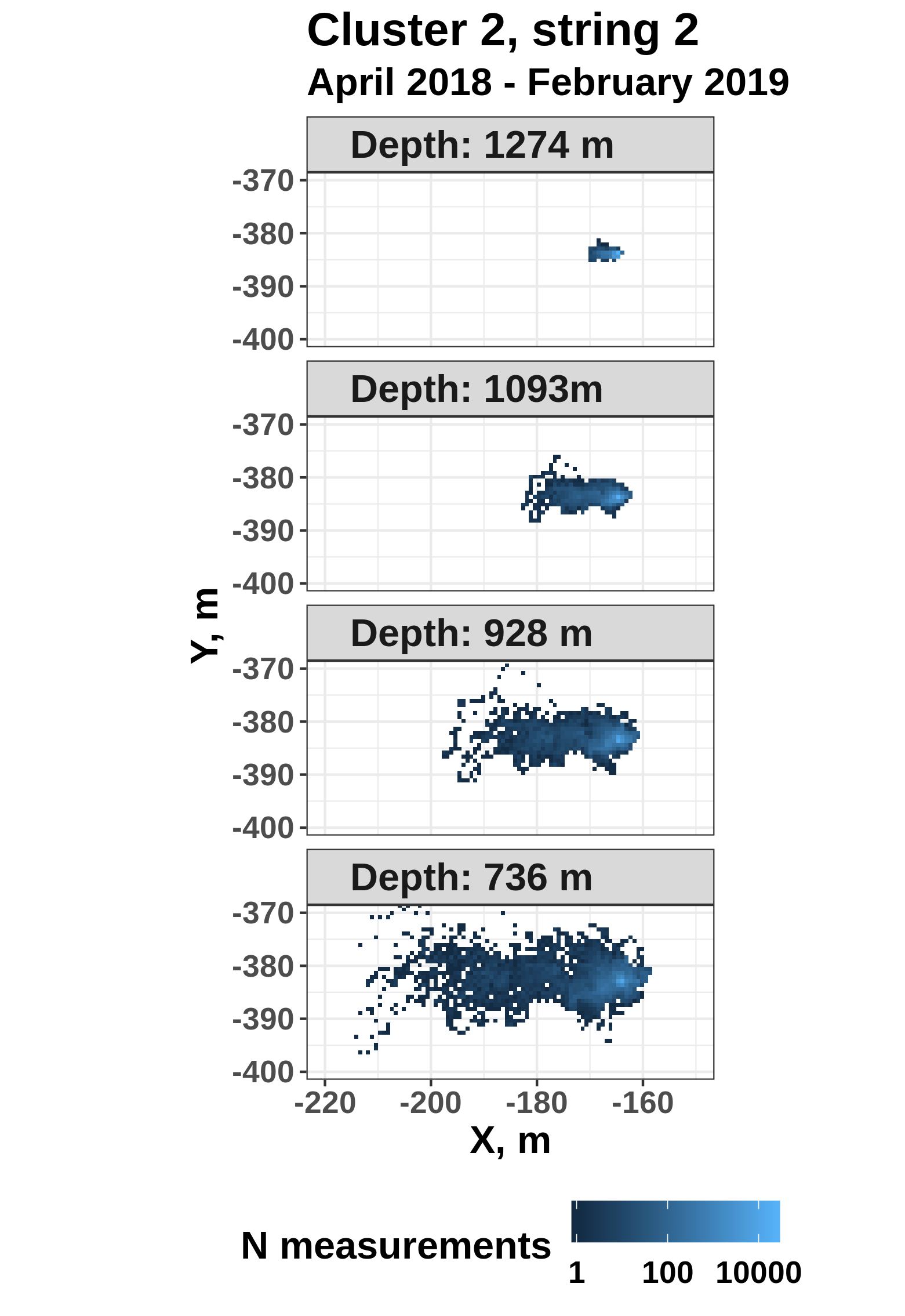}
  \end{minipage}
  \begin{minipage}{0.32\linewidth}
    \includegraphics[width=\linewidth]{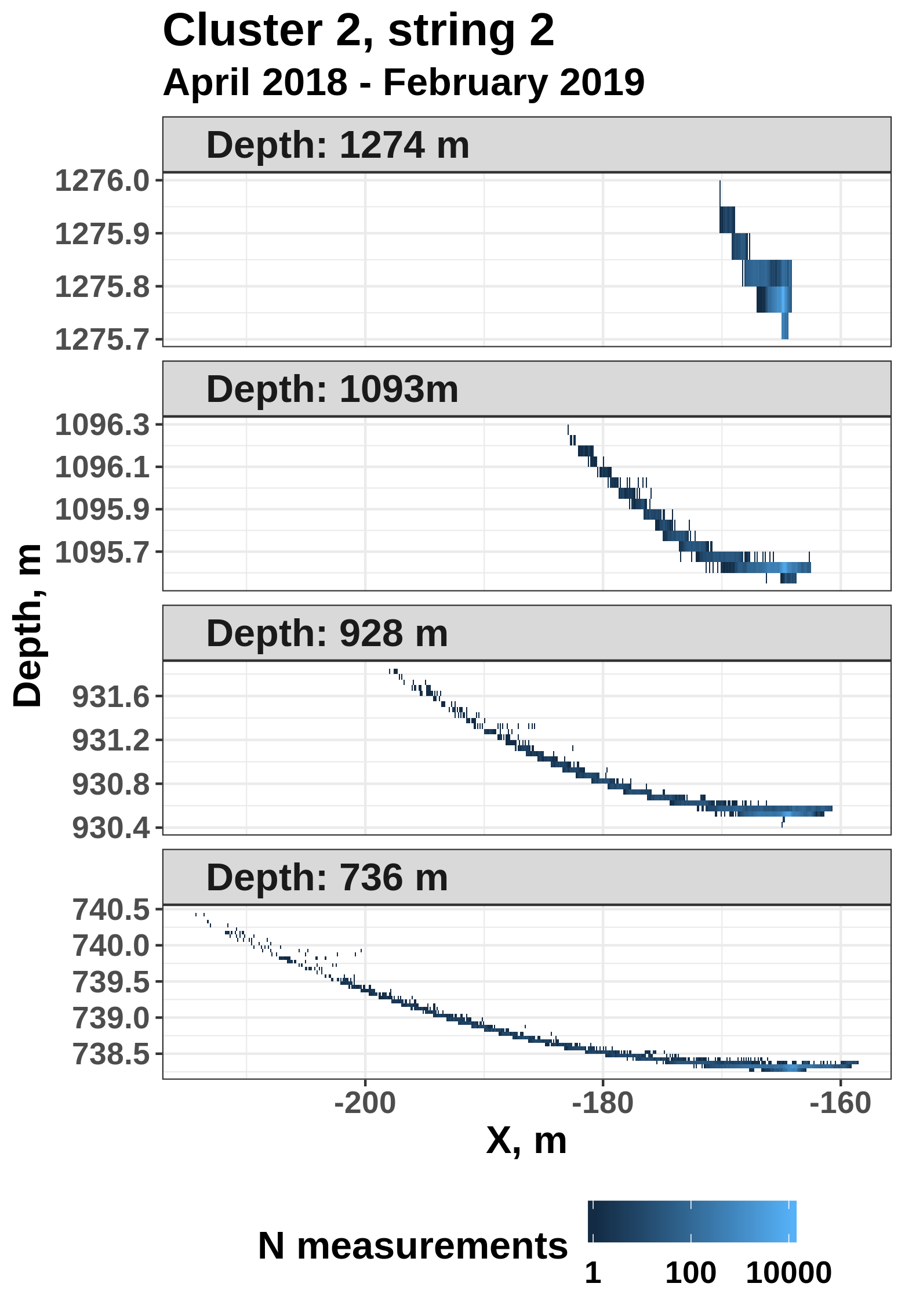}
  \end{minipage}
  \begin{minipage}{0.32\linewidth}
    \includegraphics[width=\linewidth]{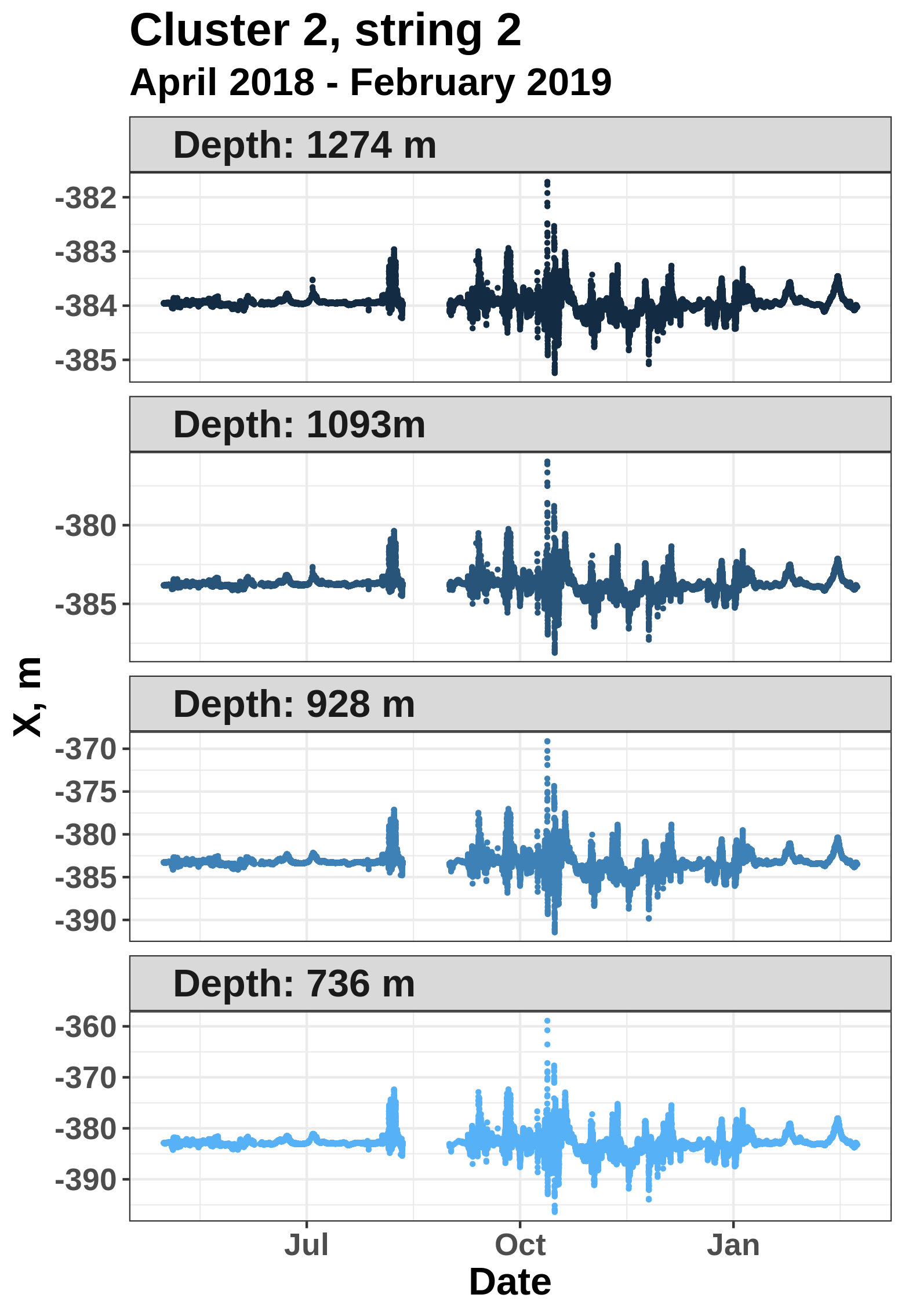}
  \end{minipage}
  \caption{Trilaterated beacon coordinates, cluster 2, string 2, season 2018}
  \label{xydrift}
\end{figure}

An example of beacon coordinate measurements for the beacons installed on one of the GVD strings is presented in Figure \ref{xydrift}. 
Beacon drift is mainly lateral, with depth variation exceeding 0.5 meters only during brief periods of active drift in autumn. 
Beacon mobility falls with depth and the coordinate variation decreases from $\sim50$ meters at the depth of 736 meters, to $\sim 5$ meters at the depth of 1274 meters. The most shallow, mobile beacons move with an average speed of 0.5 cm/s and a maximum speed of 3 cm/s.

\begin{figure}
  \begin{minipage}{0.32\linewidth}
    \includegraphics[width=\linewidth]{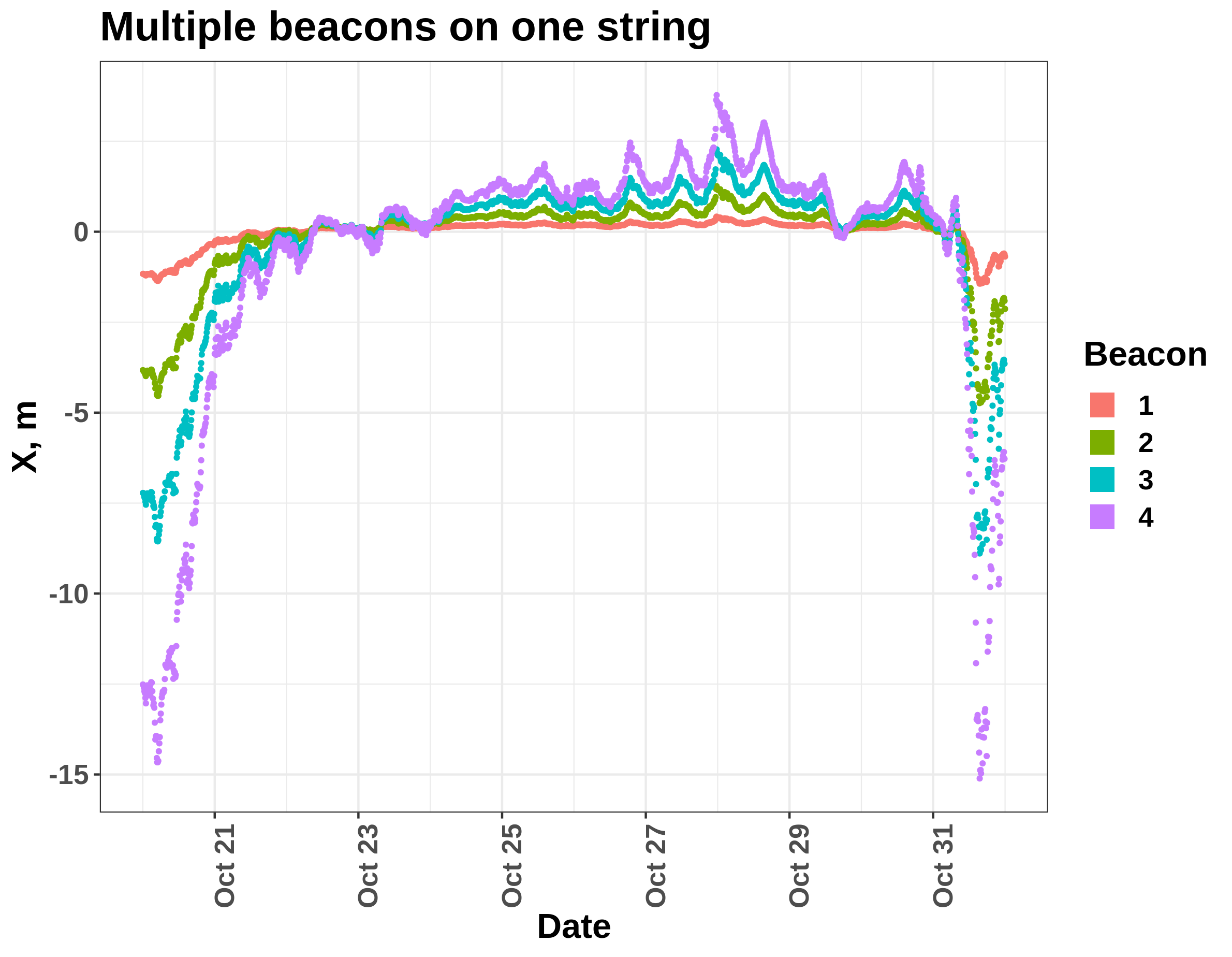}
  \end{minipage}
  \begin{minipage}{0.32\linewidth}
    \includegraphics[width=\linewidth]{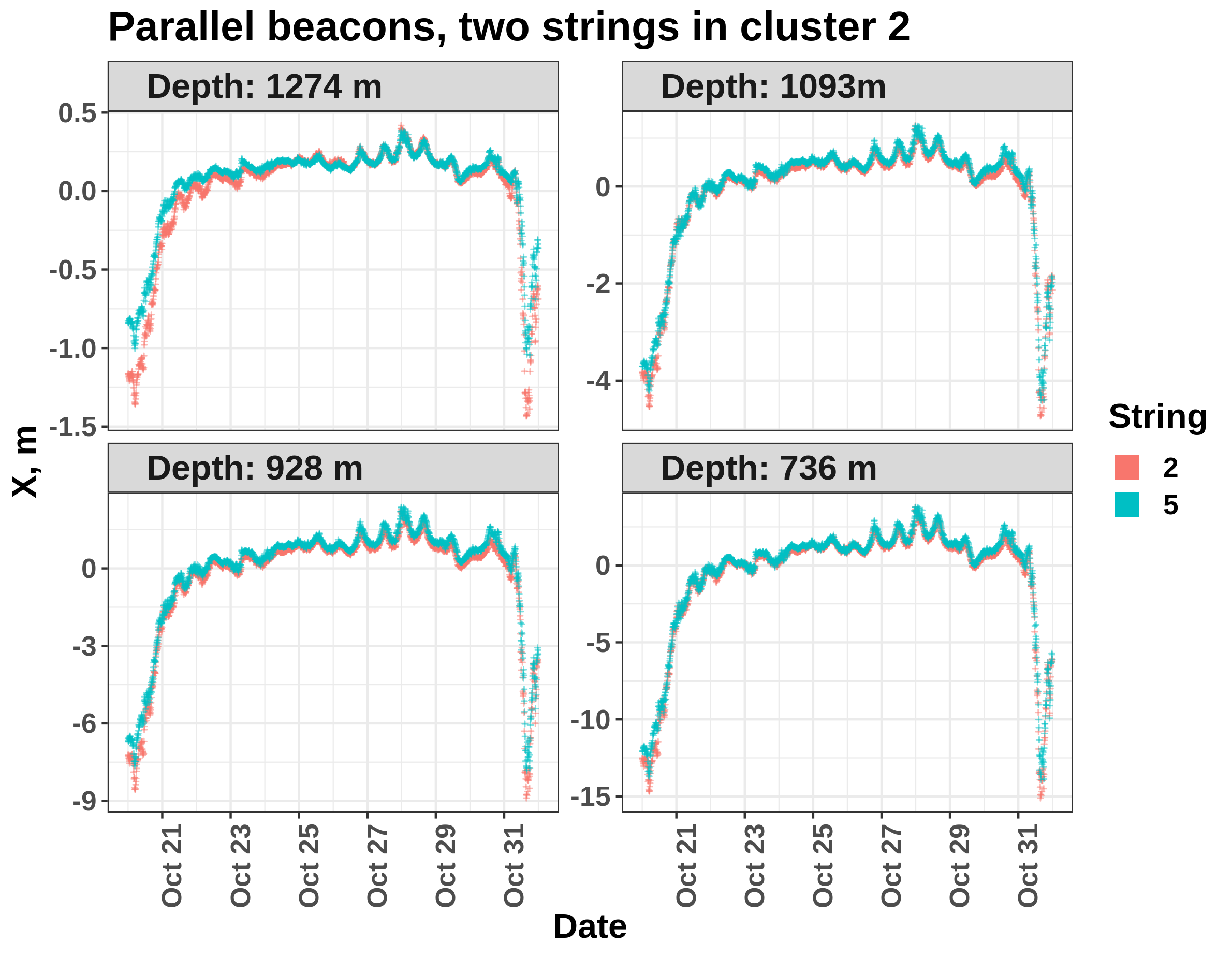}
  \end{minipage}
  \begin{minipage}{0.32\linewidth}
    \includegraphics[width=\linewidth]{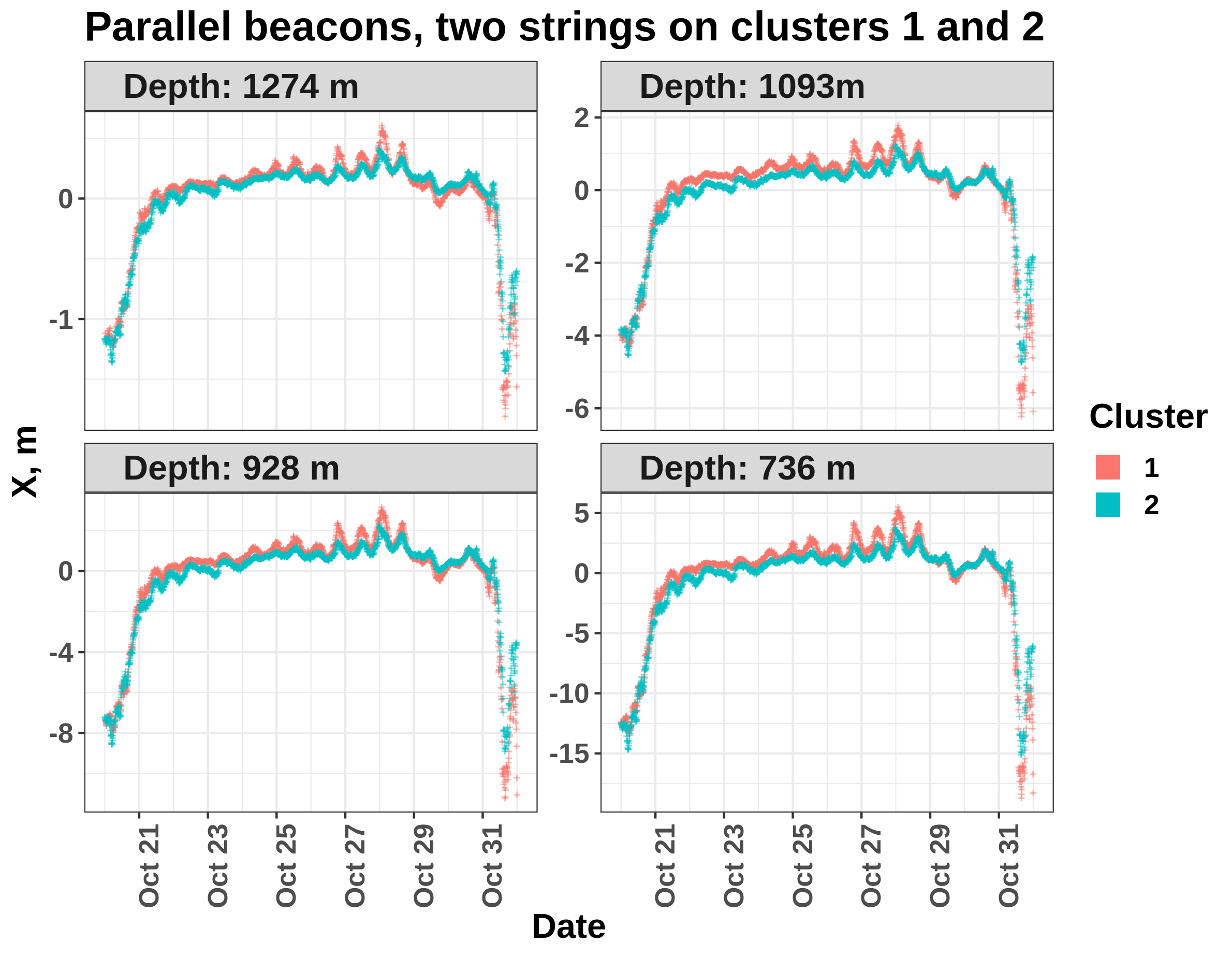}
  \end{minipage}
  \caption{Correlation between beacons, October 2018}
  \label{corr}
\end{figure}

Figure \ref{xydrift} also shows, that the coordinates of beacons installed on various depths on the same string are correlated.
This dynamic extends to beacons installed on different strings, beacons installed at the same depth within a cluster and beacons installed at the same depth on different clusters.
Figure \ref{corr} illustrates correlation between beacon coordinates from October 20th to October 31st 2018. 

\section{Positioning precision}
The OM positioning error depends on three factors.
First, the accuracy of AM measurements. 
As demonstrated in \cite{aps2011}, APS precision is within few centimeters.
Second, the error increases with the distance between the OM and the beacons used to interpolate its position.
Finally, the error depends on beacon mobility which decreases with depth and varies with season.
To provide an upper bound of the OM positioning error, additional beacons were installed on two strings of Baikal-GVD during winter expedition 2018.
The beacons were placed at the depths of 811 and 823 meters, between beacons 3 and 4 of their respective strings.
The error estimate is acquired by comparing trilaterated coordinates of these beacons with the coordinates acquired via interpolation.

\begin{figure}
  \center
    \includegraphics[height=3cm]{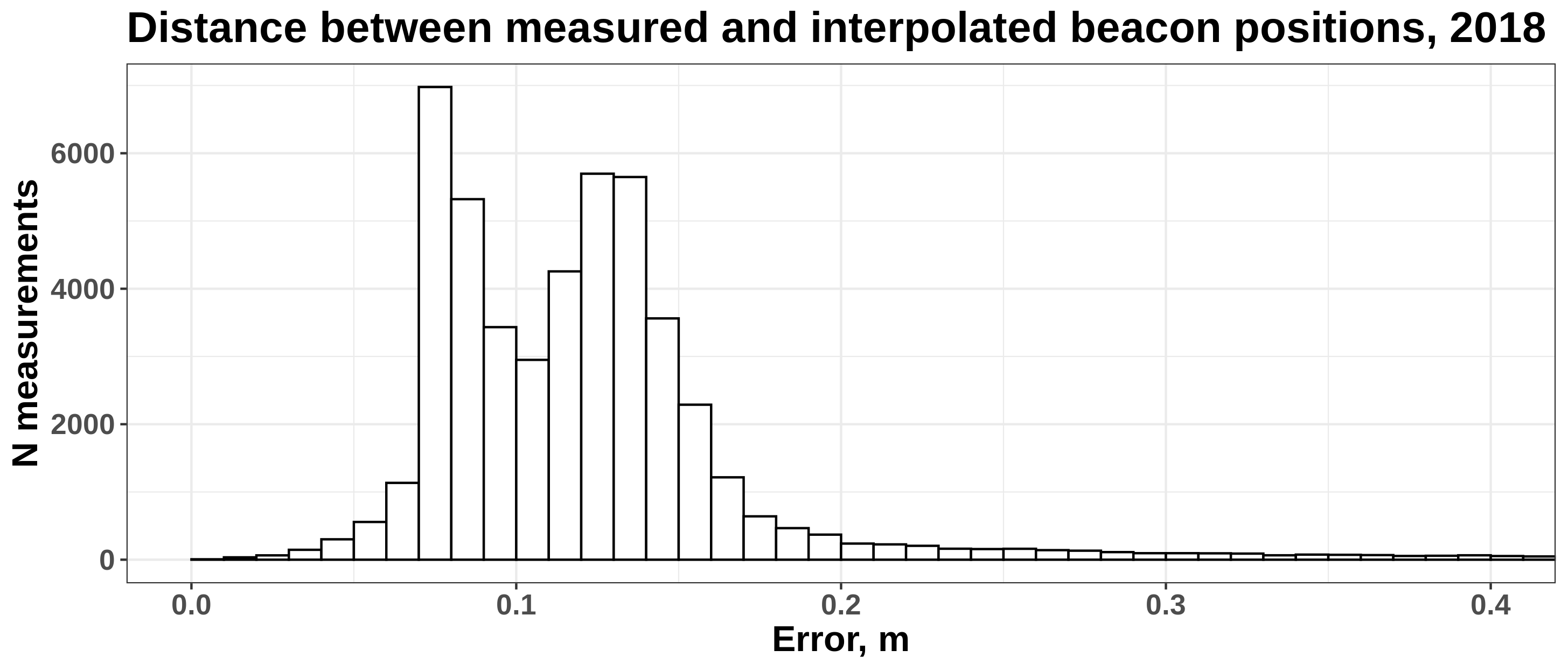}
    \caption{Positioning error estimate for April 2018 - February 2019}
    \label{precision}
\end{figure}

The results for the period from April 2018 to February 2019 are presented on Figure \ref{precision}.
The mean positioning error for both beacons over the season is 12 $\pm$ 6 cm (the photocathode diameter is 25 cm).
This is similar to the positioning precision in other large scale neutrino telescopes \cite{riccobene2019, antares2009}.

\section{Conclusions}
An acoustic positioning system has been developed and deployed at the site of the Baikal-GVD neutrino telescope.
It allows positioning optical modules of the telescope with an average accuracy of 12 $\pm$ 6 cm, which is equivalent to a subnanosecond time calibration.
Acoustic measurements have shown that over the course of a season the most mobile OMs can drift beyond 50 meters from their median positions with the average speed of 0.5 cm/s and top speed of 3 cm/s.
Acoustic data has also demonstrated that OM coordinates within a cluster and between clusters are highly correlated.

\section{Acknowledgements}
This work was supported by the Russian Foundation for Basic Research (Grants 16-29-13032, 17-0201237).



\begin{thebibliography}{99}
	\bibitem{status}
		Avrorin A. D. et al., Status, these proceedings
	\bibitem{aps2011}
    Avrorin, A. V. et al. (2013). A hydroacoustic positioning system for the experimental cluster of the cubic-kilometer-scale neutrino telescope at Lake Baikal. Instruments and Experimental Techniques, 56(4), 449-458.
	\bibitem{dmac}
    Kebkal, O. et al.  (2011, June). D-MAC: Media access control architecture for underwater acoustic sensor networks. In OCEANS 2011 IEEE-Spain (pp. 1-8). IEEE.
  \bibitem{riccobene2019}
    Riccobene, G. (2019). The Positioning system for KM3NeT. In EPJ Web Conf. (Vol. 207, p. 07005).
  \bibitem{antares2009}
    Ardid, M. (2009). Nuclear Instruments and Methods in Physics Research Section A: Accelerators, Spectrometers, Detectors and Associated Equipment, 602(1), 174-176.


\end{thebibliography}
\end{document}